\documentclass{article}

\PassOptionsToPackage{numbers, compress}{natbib}

\usepackage[final]{neurips_2024_ml4ps}

\usepackage[utf8]{inputenc} 
\usepackage[T1]{fontenc}    
\usepackage{hyperref}       
\usepackage{url}            
\usepackage{booktabs}       
\usepackage{amsfonts}       
\usepackage{nicefrac}       
\usepackage{microtype}      
\usepackage{xcolor}         
\usepackage{graphicx}
\usepackage{amsmath}
\usepackage{printlen}
\usepackage{wrapfig}
\usepackage{lipsum}
\usepackage{subcaption}
\usepackage{caption}
\usepackage{wrapfig}

\title{CASBI -- Chemical Abundance Simulation-Based Inference for Galactic Archeology}

\author{%
  Giuseppe Viterbo\\
  Interdisciplinary Center \\ for Scientific Computing\\
  University of Heidelberg\\
  D-69120 Heidelberg\\
  \texttt{giuseppe.viterbo@iwr.uni-heidelberg.de} \\
  \And
  Tobias Buck\\
  Interdisciplinary Center \\ for Scientific Computing\\
  University of Heidelberg\\
  D-69120 Heidelberg\\
  \texttt{tobias.buck@iwr.uni-heidelberg.de} \\
}

\begin{document}

\maketitle

\begin{abstract}
    Galaxies evolve hierarchically through merging with lower-mass systems and the remnants of destroyed galaxies are a key indicator of the past assembly history of our Galaxy.
    However, accurately measuring the properties of the accreted galaxies and hence unraveling the Milky Way's (MW) formation history is a challenging task.
    Here we introduce CASBI (Chemical Abundance Simulation Based Inference), a novel inference pipeline for Galactic Archeology based on Simulation-based Inference methods. CASBI leverages on the fact that there is a well defined mass-metallicity relation for galaxies and performs inference of key galaxy properties based on multi-dimensional chemical abundances of stars in the stellar halo. Hence, we recast the problem of unraveling the merger history of the MW into a SBI problem to recover the properties of the building blocks (e.g. total stellar mass and infall time) using the multi-dimensional chemical abundances of stars in the stellar halo as observable.
    With CASBI we are able to recover the full posterior probability of properties of building blocks of Milky Way like galaxies.
    We highlight CASBI's potential by inferring posteriors for the stellar masses of completely phase mixed dwarf galaxies solely from the 2d-distributions of stellar abundance in the iron vs. oxygen plane and find accurate and precise inference results.
%
\end{abstract}

\section{Motivation: The reconstruction of the assembly history of the Milky Way}
\enlargethispage{\baselineskip}
Inferring the assembly history of the Milky Way (MW) is a challenging task, even in the era of the astrometric Gaia mission and its 6 dimensional 
phase space data, and the complementary chemical information obtained from wide-field spectroscopic programs such as GALAH
\cite{desilvaGALAHSurveyScientific2015}, H3 \cite{conroyMappingStellarHalo2019}, APOGEE \cite{majewskiApachePointObservatory2017}, RAVE \cite{steinmetzRadialVelocityExperiment2006}, SEGUE \cite{yannySEGUESPECTROSCOPICSURVEY2009}, or 
LAMOST \cite{cuiLargeSkyArea2012}. While the dynamical times of the accreted objects are on the order of the age of the host galaxy, phase mixing of accreted and in-situ stars will occur and the phase space only retains part of the information on the original infall parameters. Hence, robustly identifying distinct structures is challenging, and disentangling the components in fully phase mixed situations is nearly impossible.
On the other hand, stellar chemical abundances are unchanged over the lifetime of a star serving as unique labels to tag stars. Additionally, the chemical abundance space is dependent on the star formation history and the total stellar mass of the galaxies leading to distinct differences in the abundance distribution of different galaxies \citep[e.g.][]{buck2023}.
Very recently, the crossmatch between astrometric data from Gaia with spectroscopic data allowed for the discovery of the "Gaia-Sausage-Enceladus" \citep[GSE,][]{belokurovCoformationDiscStellar2018,helmiMergerThatLed2018}, a massive accretion event whose remnant now dominates the observation of the inner stellar halo of our Galaxy. The discovery of this massive structure opened the door to unveil the merging history of our Galaxy. This work aims to exploit the information available in modern cosmological simulations to guide an automatic and data driven approach to reconstruct the properties of these objects with the use of the Simulation Based Inference \citep[SBI,][]{cranmer2020frontier}] technique.

\section{Related Work: Galactic Archaeology with Simulations}

In order to characterize the assembly history of the Milky Way (MW) from stellar chemical abundances alone \cite{cunninghamReadingCARDsImprint2022}
proposed to use the "CARDs", the chemical abundance ratio distributions of the stars, obtained from a subsample of accreted objects from the FIRE-2 zoom-in cosmological simulations of MW-mass galaxies \cite{wetzelRECONCILINGDWARFGALAXIES2016}. This method models the host halo as a linear combination of 2d templates for the joint iron and oxygen abundance:
\begin{math} 
\text{CARD}_{\text{halo, model}} (x_d) = \sum_i \sum_j A_{ij} \text{CARD}_{\text{temp}, ij} (x_d|M_{\text{sat}, i}, t_{100, j}), 
\end{math} 
treating each coefficient $A_{ij}$ as the fraction of mass contribution from the accretion event of the template satellite with mass $M_{\text{sat}, i}$ and quenching time $t_{100, j}$. CARDs tries to recover those coefficients by maximizing a loss that compares the observed abundance distribution with the combination of the templates. 
Additionally, this method does not recover full posterior for the parameters of the accreted objects but rather point estimates.
Another approach is presented in \cite{deasonUnravellingMassSpectrum2023}, which takes advantage of the mass-metallicity relation to decompose the metallicity distribution function (MDF) of the host galaxy as a mixture of accreted halo's MDFs, assuming Gaussian shape for each of these building blocks. This decomposition relies on the observational evidence that at the dwarf galaxy mass scale, not only the average metallicity varies with the mass, but also the width of the MDF \citep{kirbyMULTIELEMENTABUNDANCEMEASUREMENTS2011} where lower mass dwarfs have a wider spread of metallicities. 
This method has the fundamental problem of having a variable number of model parameters, making it difficult to sample in practice. To circumvent this, \cite{deasonUnravellingMassSpectrum2023} decided to bin the accreted dwarf galaxies in luminosity and count the number of contributions from each luminosity bin adopting a nested sampling scheme to obtain a posterior distribution for the number of galaxies in each luminosity bin.


\section{CASBI: Likelihood-free inference of MW's accretion history}
\paragraph{Simulation-based inference}
The SBI framework has existed along side the more traditional likelihood based inference methods for quite some years already, having its root in Approximate Bayes Computation \cite{rubinBayesianlyJustifiableRelevant1984}. The main difference between SBI and traditional sampling methods, like Markov Chain Monte Carlo (MCMC), is that the former do not require the likelihood function to be known, but rather rely on a simulator to generate synthetic data \textbf{$\mathbf{x}$} from input parameters $\boldsymbol{\theta}$. Inference networks are then trained based on data-parameters pairs ($\mathbf{x}, \boldsymbol{\theta}$). 
Recent advances of this technique were made possible by the use of Normalizing Flow models to emulate conditional probability distributions, a technique know as Neural Density Estimation (NDE) \cite{papamakariosNeuralDensityEstimation2019}. 
Following the discussion presented in \cite{hoLtUILIAllinOneFramework2024}, in SBI we have the choice to approximate either the Posterior, the Likelihood or the Likelihood ratio, and this choice depends mostly on the problem that one wants to solve, and in particular on the dimensionalities of \textbf{$\mathbf{x}$} and $\boldsymbol{\theta}$. In our case, due to the complexity of the Likelihood distribution of the chemical abundance space, we choose to approximate the Posterior distributions, and so we adopted the Neural Posterior Estimate (NPE). Another strong choice would have been to perform Neural Ratio Estimate (NRE), which enables one to avoid specifying an approximating distribution model and trains a classifier to target the Likelihood ratio. This kind of approach is quite capable of dealing with cases where both the Likelihood and the Posterior exhibit complicated shapes, but for our case the simple Gaussian-like shape of the posterior and the high dimensionality of the observations \textbf{$\mathbf{x}$} led to our choice of using NPE, leaving NRE for future testing. 

\paragraph{Simulator model in CASBI}
\enlargethispage{\baselineskip}
\begin{wrapfigure}[31]{R}{0.75\textwidth}
  \centering
  \vspace{-.8cm}
  \includegraphics[width=0.749\textwidth]{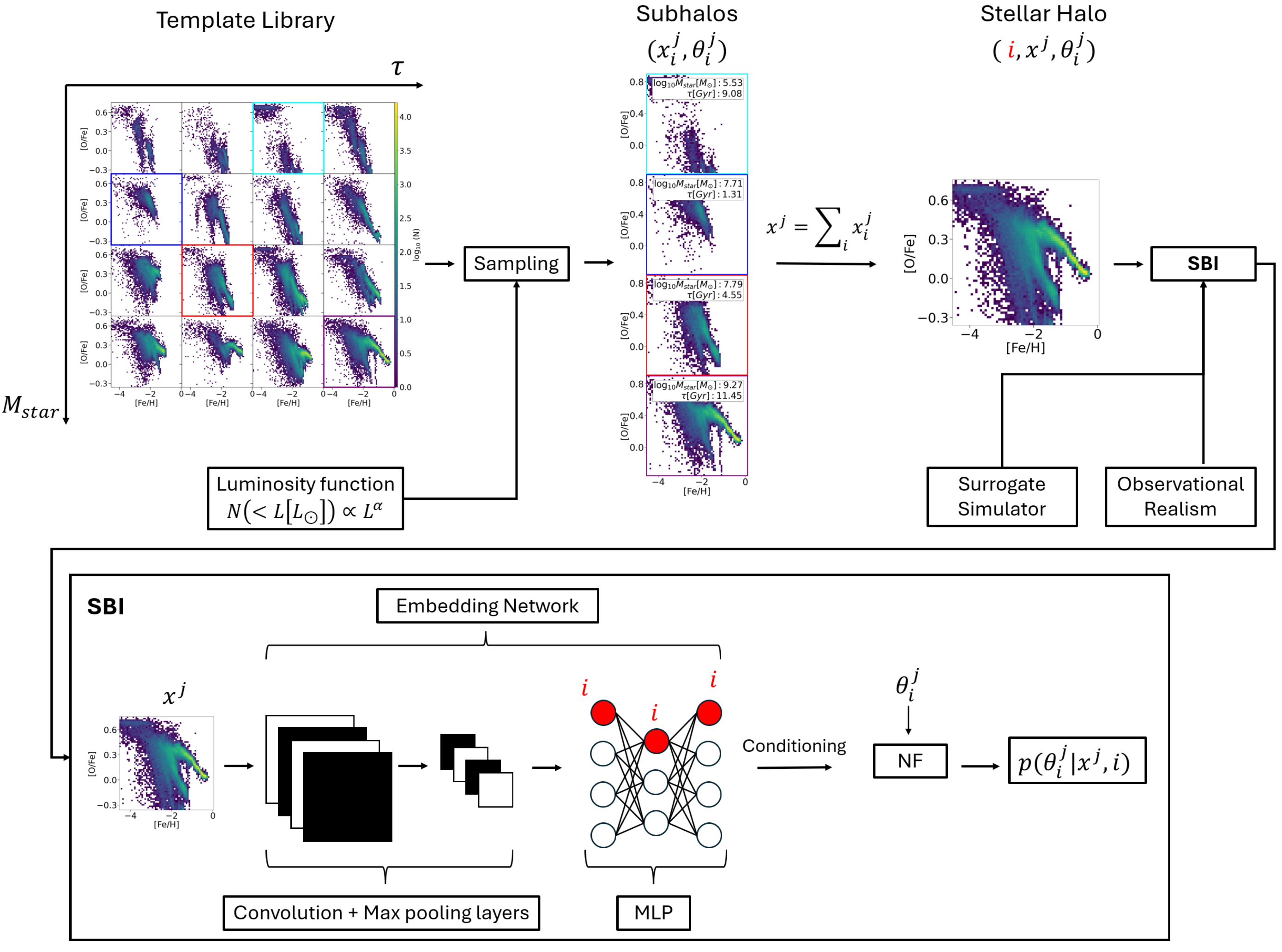}
  \caption{CASBI pipeline. In our analysis the \textbf{Template Library} is fixed to the \textbf{NIHAO} simulations. The choice of the Template Library incorporates all the assumptions that we make on the chemical enrichment histories of Galaxies, the dynamical effects that accreted objects undergo, and the cosmology, making this part the principal cause of possible misspecification. We \textbf{sample} non-repeated subhalos aiming to reproduce a \textbf{Luminosity function} $N(<L)$, \citep{koposovLuminosityFunctionMilky2008} with fixed total stellar mass budget of the halo \citep{deasonTotalStellarHalo2019}. The \textbf{SBI} pipeline can incorporate a \textbf{Surrogate Simulator} to perform the sequential version of NPE. The \textbf{Observational Realism} encapsulates all effects bridging the gap between simulation and observation (e.g. observational uncertainties, selection functions, realistic in-situ stars background, etc.).}
    \label{fig:CASBI_complete}
\end{wrapfigure}

CASBI is an SBI package to recover the properties of building blocks of MW's stellar halo from observations of the chemical abundance plane. Hence, the data $\mathbf{x}$ are multi-dimensional distributions of chemical abundances. In this proof-of-concept work we restrict ourselves to only use iron and oxygen. The parameters in CASBI ($\boldsymbol{\theta}$) can be anything describing the infalling objects, e.g. infall time, stellar mass before infall, gas mass or dark matter mass, etc.. Again, we restrict ourselves to inferring only stellar mass, $M_{star}$, and infall time, $\tau$ of the infalling dwarf galaxies for this first application, so in the end the single accreated object parameter is a vector $\boldsymbol{\theta_i} = (M_{star}, \tau)$.
As the simulator to generate the data-parameters pairs $(\mathbf{x}, \boldsymbol{\theta})$ needed to train the NPE we rely on the NIHAO cosmological hydrodynamical simulations \cite{wangNIHAOProjectReproducing2015,buckOriginChemicalBimodality2020,buck2020a,buck2021,buck2023}. \textbf{NIHAO} is a set of 100 cosmological zoom-in hydrodynamical simulations with halos that range from dwarf ($M_{star} \sim 5 \times 10^9 M_\odot$) to MW ($M_{star} \sim 2 \times 10^{12} M_\odot$) mass. 
Similar to \cite{cunninghamReadingCARDsImprint2022} and \cite{deasonUnravellingMassSpectrum2023}, we rely on the assumption that once an accreted object falls into the gravitational potential of the host galaxy its star formation is halted and its abundance distribution is frozen. This means, accreted dwarf galaxies will evolve the same as isolated dwarf galaxies prior to infall.
Under this assumption, we treat each dwarf galaxy snapshot as a possible configuration shortly before infall into the host galaxy. 
In order to create a subhalo template observation we construct 2D histograms ($64 \times 64$ pixels), referred to as \textbf{$\mathbf{x_i}$}, by binning the chemical abundance plane ([O/Fe], [Fe/H])\footnote{They are respectively proxy for $\alpha$ elements abundance and metallcity} for each of the snapshots available in \textbf{NIHAO}. We additionally filter the galaxies to only consider galaxies with a total stellar mass lower than the Large Magellanic Cloud ($M_{star} < 6 \times 10^9 M_\odot$), the largest accreted object in the MW. 
The set of all possible subhalos is defined as 'Template Library'. The actual stellar halo observable \textbf{$\mathbf{x^j} = \sum_i^{N_{sub}^j} x^j_i$ } used in CASBI is then a superposition of $N_{sub}$ of these 2D histograms, where the $N_{sub}^j$ is the number of accreted objects present in the $j$-th galaxy halo. The actual choice of how to sample from the 'Template Library' can be adapted and we use a physically informed approach by using a analytic luminosity function, taken from \cite{koposovLuminosityFunctionMilky2008} and a total stellar mass budget taken from \cite{deasonTotalStellarHalo2019}\footnote{After
obtaining the analytic samples we take the first and second Nearest Neighbors (NN) that are within 10\% of the sampled mass as our mock subhalo and reduce the total mass
budget accordingly.}. The choice of the total stellar mass budget put a constraint on the maximum subhalo mass. With this the final goal of CASBI is to recover $\mathbf{\theta^j_i}$ for each of the subhalos in the galactic halo from the observable \textbf{$\mathbf{x^j}$}$=\sum_i x^j_i$, and gaining insight on how many subhalos there are. 

\paragraph{The CASBI pipeline}
Many excellent frameworks for handling SBI analysis are already available, and CASBI is build on top of the \texttt{ltu-ili} python package \cite{hoLtUILIAllinOneFramework2024}. 
In Fig. \ref{fig:CASBI_complete} we show the CASBI pipeline. The modularity of the SBI technique is fully integrated, allowing to change all the components of this pipeline. The 'Template Library' can be adapted to any other suite of simulated galaxies (e.g. \cite{pillepichMilkyWayAndromeda2023}), the sampling scheme can incorporate different luminosity functions and stellar halo budgets, the NPE and embedding network architecture and hyper-parameters can be modified to allow for higher accuracy and posterior coverage thanks to the \texttt{optuna} grid search implementation, and surrogate models\footnote{We leave the integration of the surrogate model for future work} (e.g. Free Form Flow FFF \cite{draxlerFreeformFlowsMake2024} surrogates, or semi-analytical galaxy formation models such as GRUMPY \cite{kravtsovGRUMPYSimpleFramework2022}) can be implemented to allow for the sequential version of the NPE.

\enlargethispage{\baselineskip}
\paragraph{Final CASBI model architecture and training data}
\label{sec: model architecture and training data}
After an extensive hyperparameter tuning and model evaluation phase using \texttt{optuna} Multi-objective Optimization, we have found that the 'Neural Spline Flow' (nsf) model  with 100 hidden units and 20 transformations available in the \texttt{lampe} back-end  of \texttt{ltu-ili}, achieves both the highest calibration and log posterior. As described in \cite{hoLtUILIAllinOneFramework2024}, a single NDE tends to be overconfident, so it underestimates uncertainty; in order to deal with this problem one can ensamble multiple models and take a weighted sampling scheme accordingly to their Posterior value. In our case we ensambled four nsf models to obtain our final results. The embedding network is a Convolutional Neural Network (CNN) with 3 convolutional, 3 Max Pooling and 4 fully connected layers with a final embedding dimension of 32 which was obtained during the hyperparameter tuning phase. We have generated 1000 training set stellar halo and 100 test set stellar halo with a maximum of 100 subhalos each. The training took $\sim$ 1 hour on a single NVIDIA A100.  

\section{Results}
\label{sec: Results}
\vspace{-1em}
\begin{wrapfigure}[15]{R}{0.575\textwidth}
    \centering
    \vspace{-0.75cm}
    \includegraphics[width=0.57\textwidth]{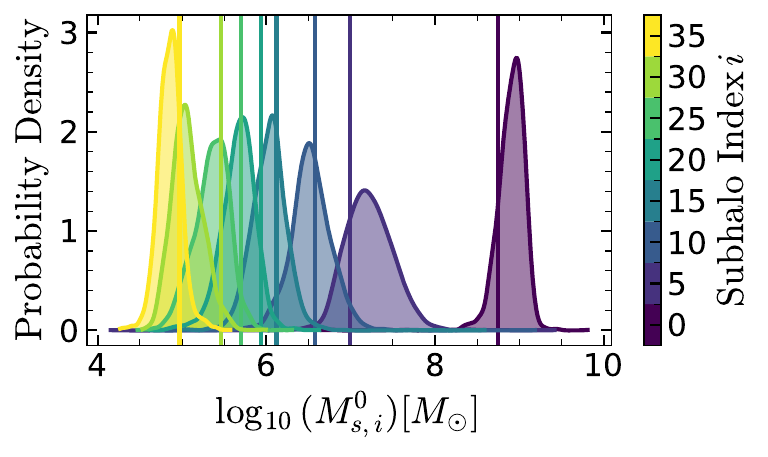}
    \vspace{-.8cm}
    \caption{Posterior of the stellar mass $M_s$ for subhalos with index $i$ belonging to the first stellar halo of the test set ($j=0$). The vertical lines show the true values for each subhalo.}
    \label{fig:Posterior}
\end{wrapfigure}
\begin{figure}[htbp]
    \centering
        \begin{subfigure}[c]{.32\textwidth}
            \centering
            \includegraphics[width=\textwidth, clip]{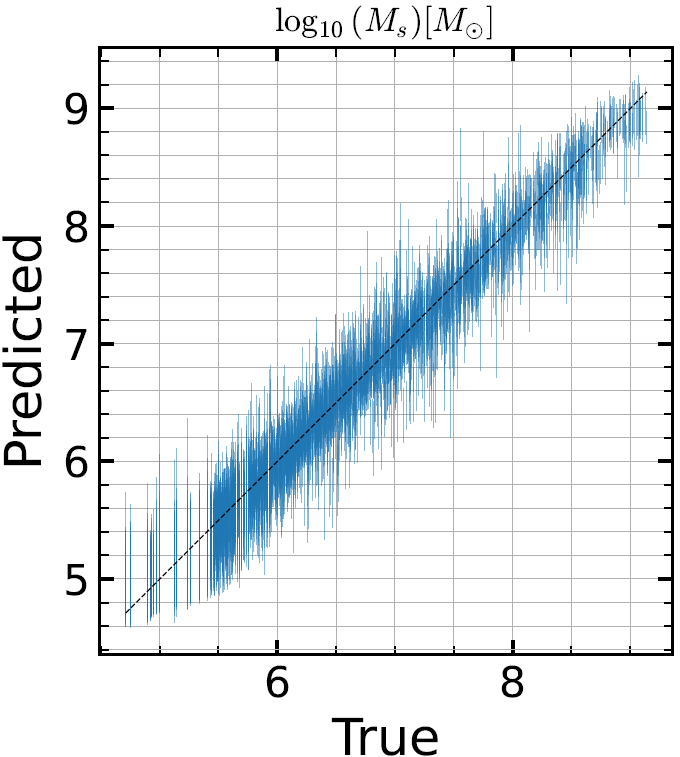}
            \caption{True and Predicted }
            \label{fig:True_vs_Predicted}
        \end{subfigure}
        \begin{subfigure}[c]{.32\textwidth}
            \centering
            \includegraphics[width=\textwidth,trim={0cm 0cm 9.15cm 0cm},clip]{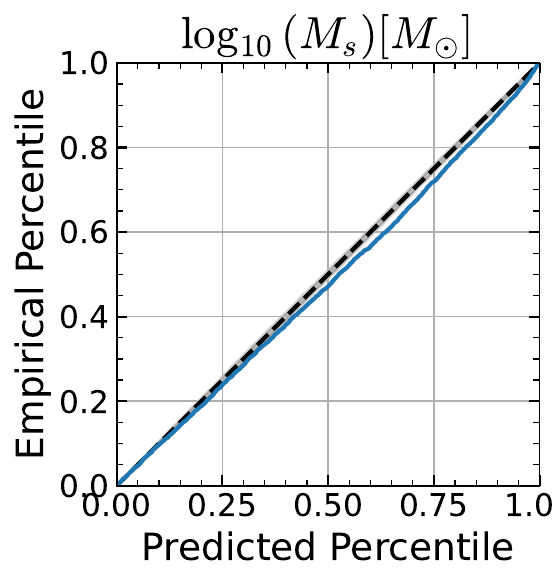}
            \caption{P-P plots }
            \label{fig:ppplots}
        \end{subfigure}
        \begin{subfigure}[c]{0.335\textwidth}
            \centering
            \includegraphics[width=\textwidth,trim={0cm 0.35cm 0cm 0cm},clip]{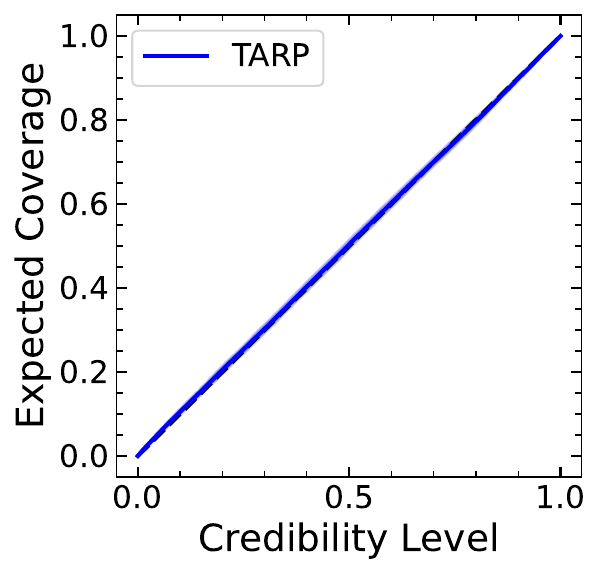}
            \caption{TARP}
            \label{fig:Tarp}
        \end{subfigure}
    \caption{Inference results on the test set. We show the True vs Prediced values for the stellar mass $M_{s}$ part of the $\boldsymbol{\theta}$, with error bars representing the 68\% confidence interval in (a), the marginal posterior coverage  for the stellar mass $M_{s}$ in (b), and full posterior coverage in (c).}
    \label{fig:Inference Results}
\end{figure}

 
In Fig.~\ref{fig:Posterior} we present the posterior probability of subhalo masses for a test set stellar halo. In this example we can appreciate that the true value for each of the subhalos falls close to the posterior estimate (i.e. the posterior mean) and that posterior predictions are relatively peaked around the true
values. This finding makes us confident that CASBI is able to infer infall properties of completely phase mixed dwarf galaxies solely from chemical abundances with relatively high precision and accuracy. In order to correctly evaluate the constraining power and the calibration of the model we report in Fig.~\ref{fig:Inference Results} the comparison between true and predicted values of the parameters in the test simulations, the coverage tests using both the percentile-percentile (P-P) plots and the TARP test \cite{lemosSamplingBasedAccuracyTesting}\footnote{TARP constructs, in the limit of sufficient samples, an estimate of the untractable joint posterior coverage which is guaranteed to converge to the true posterior coverage. In our case, since the infall time $\tau$ component of $\boldsymbol{\theta}$ is not constrained and its posterior resamples the prior distribution, the TARP value is less meaningful than the marginal coverage plot}. Looking at \ref{fig:True_vs_Predicted}, we find good predictive performance on the stellar mass $M_s$ across the entire mass range probed. Similarly, from Fig.~\ref{fig:ppplots} and Fig.~ \ref{fig:Tarp} we see that the model results appear to be well calibrated both in the marginal and in the full posterior.
We have further tested CASBI's ability to infer infall times, $\tau$, from chemical abundances alone but found poor performance for this task. The lack of predictability of the infall times can be associated to the degeneracy of the chemical abundances between a massive system that was accreted early on and a less massive system that was accreted more recently. The integration of the orbital information or using more $\alpha$ elements (e.g. Mg, Si, Mn) could alleviate this limitation of the model. Both of this integration are left as future work. 
\vspace{-1.15em}
\vspace{-2em}
\enlargethispage{\baselineskip}
\section{Conclusion and limitations}
\label{sec: Conclusion and Limitation}
\vspace{-.5em}
We find that SBI can be applied to the complex task of parameter estimation for accreted dwarf galaxies in the MW, leveraging only on the chemical information. This allows for the inference of infall properties even in the case of fully phase mixed objects in the Galactic halo and opens up a completely new avenue for Galactic Archeology. We find that the information of multi-dimensional chemical abundance distributions is able to guide the model towards a well calibrated and accurate reconstruction of the stellar mass of accreted dwarf galaxies. Furthermore, our method avoids the need 
to bin the subhalo information as was done in \cite{deasonUnravellingMassSpectrum2023}.
In this way with CASBI it is possible to directly perform Bayesian inference on the parameters of each individual subhalo. 
One limitation of our method is the simplified scenario in which we have tested CASBI. We are asuming to be able to perfectly remove background stars and perform inference on a clean halo sample of stars. Additionally, we have not yet assumed any observational selection function nor observational uncertainties on stellar abundances. But all this will be implemented in the 'Observational Realism' component of CASBI and will be evaluated in future work (see Fig.~\ref{fig:CASBI_complete}).
%
Our code for CASBI including extensive documentation is available on \href{https://github.com/ca900/ml4science_casbi/blob/main/paper/paper_inference.ipynb}{GitHub}  as well as the \href{https://doi.org/10.5281/zenodo.13730400}{data} to reproduce the results of this paper. 

\section*{Broader impact statement}
The authors are not aware of any immediate ethical or societal implications of this work. This work purely aims to aid scientific research and proposes to apply novel SBI techniques to unravel the formation history of the Milky Way.

\begin{ack}
The authors thank the Scientific Software Center at Heidelberg University for the support. This work is funded by the Carl-Zeiss-Stiftung through the NEXUS programm.
\end{ack}

\section*{References}

{
\small
\bibliographystyle{plain}
\bibliography{bib.bib}
}

\end{document}